# Phase growth control in low temperature PLD Co:TiO$_2$ films by pressure

S. Rout [a], N. Popovici [a], S. Dalui [a,b], M.L. Paramês [a], R.C. da Silva [c], A.J. Silvestre [b], O. Conde [a,*]

[a] *University of Lisbon, Department of Physics and ICEMS, 1749-016 Lisboa, Portugal*
[b] *Instituto Superior de Engenharia de Lisboa and ICEMS, 1959-007 Lisboa, Portugal*
[c] *Instituto Superior Técnico, ITN, Department of Physics, 2686-953 Sacavém, Portugal*

---

* Corresponding author:

omconde@fc.ul.pt  (Olinda Conde)

Tel: +351 21 7500035  ;  Fax: + 351 21 7500977

# Phase growth control in low temperature PLD Co:TiO$_2$ films by pressure

S. Rout [a], N. Popovici [a], S. Dalui [a,b], M.L. Paramês [a], R.C. da Silva [c], A.J. Silvestre [b], O. Conde [a,*]

[a] *University of Lisbon, Department of Physics and ICEMS, 1749-016 Lisboa, Portugal*
[b] *Instituto Superior de Engenharia de Lisboa and ICEMS, 1959-007 Lisboa, Portugal*
[c] *Instituto Superior Técnico, ITN, Department of Physics, 2686-953 Sacavém, Portugal*

**ABSTRACT**

This paper reports on the structural and optical properties of Co-doped TiO$_2$ thin films grown onto (0001)Al$_2$O$_3$ substrates by non-reactive pulsed laser deposition (PLD) using argon as buffer gas. It is shown that by keeping constant the substrate temperature at as low as 310 ºC and varying only the background gas pressure between 7 Pa and 70 Pa, it is possible to grow either epitaxial rutile or pure anatase thin films, as well as films with a mixture of both polymorphs. The optical band gaps of the films are red shifted in comparison with the values usually reported for undoped TiO$_2$, which is consistent with *n*-type doping of the TiO$_2$ matrix. Such band gap red shift brings the absorption edge of the Co-doped TiO$_2$ films into the visible region, which might favour their photocatalytic activity. Furthermore, the band gap red shift depends on the films' phase composition, increasing with the increase of the Urbach energy for increasing rutile content.

## 1. Introduction

Titanium dioxide (TiO$_2$) has emerged as an extremely valuable transition metal oxide with numerous applications in a variety of technologically important areas due to its unique electronic and optical properties. TiO$_2$ has been extensively investigated as a promising photocatalyst for the treatment of industrial wastewaters and polluted air since it has a good photoactivity under ultraviolet light irradiation, it is non-toxic, water insoluble, and inexpensive [1,2]. Moreover, TiO$_2$ remains one of the best candidate materials in the development of dye solar cells due to its high efficiency, chemical inertness and photostability [3]. The major obstacle for its wide practical application in photocatalysis and photoelectrochemical energy conversion is its wide band gap that limits the photogeneration of electrons and holes to the UV region (comprising only ~5% of the solar radiation reaching the Earth). Therefore, much effort has been concentrated on the narrowing of the optical band gap of titania, namely by doping with transition metals such as Fe [4], Cr [5] and V [6]. On the other hand, Co-doped TiO$_2$ in anatase and rutile forms has gained immense interest, being one of the first oxide-based diluted magnetic semiconductor (DMS) systems to be investigated [7,8], in parallel with Co:ZnO [9], where a Curie temperature well above room temperature ($T_C > 650$ K) was demonstrated [7,8,10,11].

There are several experimental reports on the growth of either anatase or rutile phase of Co:TiO$_2$ depending on the deposition technique, growth conditions and substrates used. Yang *et al*. [12] have shown that films grown onto (100)Si and (0001)Al$_2$O$_3$ substrates always exhibit the rutile phase unless a buffer layer is used, e.g. SrTiO$_3$/TiN, in between the substrate and the film. On the other hand, the use of (001)LaAlO$_3$ substrates yields the growth of anatase thin films, but not rutile, independently of the deposition technique, which can be explained by the very small lattice misfit between (001)LaAlO$_3$ substrate and the TiO$_2$ anatase phase [13]. Shinde *et al*. [14]

have deposited anatase Co:TiO$_2$ films onto SrTiO$_3$ by pulsed laser deposition (PLD) at a substrate temperature of 700 °C whereas Chambers *et al*. [15] have shown the non-dependence of anatase phase growth on substrate temperature.

Previously, we have studied the magnetic properties of Co:TiO$_2$ films prepared by PLD in a reducing atmosphere containing hydrogen [16,17]. In this work, we focus on the possibility to control phase growth and band gap by simply and neatly varying just one experimental parameter, the argon background pressure in non-reactive PLD experiments. We show that the variation of the argon gas pressure by one order of magnitude is sufficient for the as-grown films on c-cut sapphire to evolve from an epilayer of rutile to a pure anatase thin film. Furthermore, the films were grown at a relatively low temperature, 310 °C, compared with the deposition and/or annealing temperatures usually reported [13,14].

## 2. Experimental

Co-doped TiO$_2$ thin films were deposited onto (0001)-oriented Al$_2$O$_3$ single crystal substrates by pulsed laser deposition (PLD) using a KrF excimer laser, with $\lambda$ = 248 nm and pulse duration of 30 ns, operated at a repetition rate of 20 Hz. Before deposition the substrates were ultrasonically cleaned in acetone and rinsed with deionized water. A ceramic target consisting of rutile TiO$_2$ with addition of 1 mol% Co$_3$O$_4$ was used for the deposition of the thin films. The target-to-substrate distance was kept constant at 45 mm. In order to minimize the interaction between the laser radiation and the plasma plume formed during the deposition, the laser beam was incident at an angle of 45° with respect to the target surface. Prior to any experiment, the PLD chamber was evacuated to a base pressure of 10$^{-4}$ Pa and the laser beam was rastered over the target surface for cleaning. During ablation, the target was continuously rotated and periodically translated to avoid the development of surface microstructures which might have deleterious effects on the chemical composition of the growing films. Argon was used as background gas and no supplementary source of oxygen was employed. All experiments were carried out keeping constant the laser fluence, the substrate temperature and the argon flow rate at 6.4 Jcm$^{-2}$, 310 ºC and 30 sccm, respectively. The working pressure, $P_T$, was varied between 7 and 70 Pa. After deposition, the films were slowly cooled to RT by turning off the power at the substrate heater, maintaining the ambient gas in the chamber.

The as-grown thin films were investigated by a number of complementary characterization techniques. X-ray diffraction (XRD) was used for phase analysis and crystallinity studies of the films. The diffraction patterns were recorded in the $\theta$–$2\theta$ coupled mode using CuKα radiation, a step size of 0.02º, and time per step of 15 s. The samples' chemical composition and homogeneity were evaluated by Rutherford backscattering spectrometry (RBS), which was performed using a 2.0 MeV He$^+$ beam. The backscattered ions were detected by means of two silicon surface barrier detectors, placed at angles of 140º and close to 180º to the beam direction, and with energy resolutions of 13 keV and 18 keV, respectively. The thickness and the optical band gap of the films were determined from optical transmittance measurements carried out in a UV–Vis–NIR spectrophotometer (300 nm – 1100 nm) with a bare substrate placed in the reference beam path. The deposition rate, $r_{dep}$, was obtained by dividing the thickness of a film by the number of laser pulses used for growing that particular film. The microstructure of the films was examined with a scanning electron microscope equipped with a field emission gun (FEG-SEM). For comparison purposes, some of the samples were cut transversally and their cross-section observed with the FEG-SEM for thickness estimation.

3. Results and discussion

The chemical composition of the samples was evaluated by Rutherford backscattering analyses. Fig. 1 shows a typical RBS spectrum, obtained for a film deposited at $P_T = 10$ Pa. The depth profile of the elements was extracted from the data using the RUMP simulation program [18], and looking for the closest match with the experimental data. The Ti profile shows an extended plateau indicating a homogeneous distribution along the film depth. Cobalt gives a weak but detectable contribution to the RBS spectra due to its presence in low concentration, as shown by the small step at the foot of the Ti edge in the RBS spectrum. The agreement between the experimental data and the simulated profile was clearly improved by setting an homogeneous distribution of Co along the film depth. A Ti:Co ratio of 0.97:0.03 was estimated from the RBS data, similar to the nominal composition of the target material. Moreover, similar chemical compositions were obtained for all the films investigated in this work.

The XRD pattern of a film prepared at the lowest background gas pressure of 7 Pa is displayed in Fig. 2a. The diffraction peaks were indexed on the basis of the rutile and anatase phases, using the JCPDS database files 21-1272 and 21-1276, respectively. Besides the characteristic peaks of the c-cut sapphire substrate, at $2\theta = 41.72°$ and $90.76°$, only two other peaks were found, at $2\theta = 39.47°$ and $84.91°$, which can be attributed to the (200) and (400) lattice planes of rutile. The film is thus highly oriented to the substrate. In order to examine the epitaxial growth of the (100) Co-doped $TiO_2$ film on the (0001) sapphire substrate and their relative orientation, a scan along the film $\varphi$–axis was recorded. The scan was performed around the asymmetrical $[10\bar{1}8]$ crystallographic plane of sapphire ($2\theta = 61.3°$), which makes a $\psi$ tilt angle of $21.51°$ with the basal plane. The film $\varphi$–scan was made around the (301) plane of rutile ($2\theta = 69.01°$) for which $\psi = 26.80°$. Fig. 2b shows the $\varphi$–scan plots recorded between 0° and 360°. For sapphire, the expected three-fold symmetry of the basal plane is observed while for the rutile film the $\varphi$–scan plot displays six reflection peaks separated by 60º, indicating that the film contains three domains rotated by 120° from each other, which can be understood if one considers a plane of two-fold symmetry such as the rutile (100) on a three-fold symmetry surface [19]. Furthermore, the agreement of the peaks location between the two plots shows that the basal [001] direction of Co-doped $TiO_2$ and the [10$\bar{1}$0] direction of sapphire are coincident. It is worth noting that the unequal intensities observed for symmetry related peaks are due to the fact that the scans were performed using a conventional diffractometer without sample tilt perpendicular to the beam. The inset of Fig. 2a shows the rocking curve (RC) of the (200) reflection for the rutile film. Its full-width at half-maximum (FWHM) is 1.26º while the RC-FWHM of the (0001)$Al_2O_3$ substrate was measured as 0.19º. This means that, despite the epitaxial growth of the rutile $Co:TiO_2$ film at 7 Pa, it presents some crystallite mosaic spread which might be due to the lattice mismatch of –3.7% between the (100) rutile plane and the sapphire basal plane.

The diffractograms of the films grown at higher pressures, $P_T$ > 7 Pa, are very different from the previous one and, therefore, we have plotted them in a separate figure. Fig. 3 shows the XRD patterns of the films obtained at gas pressures in the range 10 – 70 Pa, where only the relevant regions are shown i.e., the regions where peaks assigned to the $Co:TiO_2$ phase do exist. As can be observed, by increasing the working pressure up to 20 Pa the $Co:TiO_2$ films become polycrystalline, the rutile structure displaying new diffraction lines with the (101) lattice plane acquiring some prominence. At a pressure of 30 Pa the anatase phase starts to grow, and between 30 Pa and 40 Pa the diffraction peaks of anatase and rutile display similar intensities. At higher pressures the anatase phase is prevalent. No traces of secondary phases in the Ti-O-Co composition triangle were detected in the XRD patterns of the samples prepared throughout the whole range of $P_T$ values investigated, even when the diffracted intensity was plotted on a logarithmic scale (not shown). Therefore, the XRD analyses seem to provide evidence for the homogeneous distribution of the doping element, within the detection limit of the technique, in accordance with the RBS analyses.

For the polycrystalline films exhibiting a mixture of both phases, 30 Pa $\leq P_T \leq$ 50 Pa, the relative volume fraction of nanocrystalline anatase, $C_A$, and nanocrystalline rutile, $C_R$, present in the films was estimated using the direct comparison method described by Cullity and Stock [20]:

$$\frac{C_A}{C_R} = \frac{R_R I_A}{R_A I_R} \tag{1}$$

where $I_A$ and $I_R$ are the integrated intensities of the (101) anatase and (101) rutile peaks, and $R_i = LP_i |F_i|^2 p_i / v_i^2$ ($i = A,R$) where $LP_i$, $F_i$, and $p_i$ are the combined Lorentz-polarization factor, the structure factor and the multiplicity of the considered (hkl) reflection, respectively, and $v_i$ the unit cell volume of each phase. It was assumed that the temperature factor is the same for both phases. Outside the aforementioned pressure range, $C_A$ or $C_R$ were taken as 100%. Fig. 4 presents the $C_A$ and $C_R$ values as a function of the total pressure, where three distinct regions can be clearly identified: a region of pure rutile phase at lower pressure, $P_T \leq 20$ Pa, a region of rutile-anatase phase mixture, 20 Pa $< P_T <$ 60 Pa, and a region of pure anatase phase, $P_T \geq 60$ Pa. The increase of $C_A$ with $P_T$ is initially slow, stepping up before slowing down again when approaching the upper limit $C_A = 1$. The $C_A(P_T)$ data follows a pattern that resembles a sigmoid function, however the point of inflection is clearly located in the early part of the curve which is asymmetrical about the inflection. This is well described by a Gompertz-like sigmoid function [21] as follows:

$$C_A = a\,e^{-e^{-k(P_T - b)}} \tag{2}$$

with fitting parameters $a = 1.01 \pm 0.01$, $k = 0.13 \pm 0.78$ Pa$^{-1}$ and $b = 34.94 \pm 0.32$ Pa, a correlation coefficient $R^2$ = 0.999, and the inflection point located at $P_T$ = 34.94 Pa and $C_A$ = 0.373.

The influence of $P_T$ on the relative amount of rutile and anatase phases in the films may be understood by considering the interactions between the ablated species and the background gas. In the case of the films grown at the lowest pressures, the interaction between the ablated species and the background gas is weaker. Energetic particles land on the substrate transferring their kinetic energy to the substrate surface and/or growing film. As a consequence, surface migration is enhanced favoring the formation of the phase that has the most compact structure, rutile. By increasing $P_T$, the kinetic energy of the ablated species is reduced due to the collisions with the background argon gas, thus favoring the more open structure (anatase). On the other hand, in order to increase the total pressure keeping the Ar flow rate constant, the vacuum system extraction capacity should decrease and, consequently, the oxygen partial pressure, $P_O$, increases. A higher $P_O$ may be at the origin of an excess of oxygen in the films yielding the preferential growth of the anatase phase due to its lower density as compared to rutile. These results are consistent with those (not shown) obtained in experiments performed at a constant pressure of 50 Pa and varying the Ar flux from 6 sccm to 250 sccm: as the Ar flow rate decreases, implying that the partial pressure of oxygen increases, the anatase content in the films also increases. Besides the rutile/anatase evolution with increasing pressure, it is also interesting to note that the inflection point on the Gompertz curve corresponds to the maximum value of the films deposition rate, $r_{dep}$. As seen in the inset of Fig. 4, the variation of the deposition rate with background pressure follows the derivative curve of the Gompertz function, attaining a maximum value located close to the inflection point. At a first glance one would be tempted to fit the $r_{dep}$ data points with a single function. However, this is not the correct procedure because the right and left extreme data points correspond to different phases. Therefore, the slight increase of the deposition rate with pressure on the rutile side may be attributed to the change of film growth regime yielding epilayer growth at 7 Pa but polycrystalline layers at 10 and 20 Pa. On the contrary, we interpret the decrease of the deposition rate with increasing pressure for the anatase containing films as a result of the multiple elastic scattering undergone by the ablated particles due to the background gas, in agreement with the results of Amoruso *et al.* [22] obtained for PLD oxide films under similar pressure conditions.

The average crystallite size of the films along the [$hkl$] direction, $<D>_{hkl}$, was estimated from the broadening of the Bragg reflection peaks measured at half their maximum intensity, $B$, using Scherrer's equation: $<D>_{hkl} = K\lambda / B\cos\theta$, where $K$ is a constant (typically $K\sim0.9$) [23]. Fig. 5 shows the $<D>_{hkl}$ values as a function of $P_T$, which were calculated from the (200) diffraction line of the epitaxial rutile film prepared at $P_T = 7$ Pa and the (101) diffraction line either for the polycrystalline rutile films or for the prevalent anatase ones. As a general trend, it can be seen that the mean crystallite size decreases for both phases as the working pressure increases i.e., the films become less crystalline with increasing $P_T$.

Fig. 6 displays the SEM micrographs of films showing two different phase compositions and the corresponding particle size distribution histograms. The highly oriented rutile film prepared at $P_T = 7$ Pa (Fig. 6a) shows a very uniform and dense microstructure, with a *quasi* lognormal particle distribution with a mean value of 46±14 nm, in reasonable agreement with the value determined from the XRD pattern of this sample. By increasing the total pressure, the anatase content increases and the microstructure of the films changes noticeably. Fig. 6b shows the microstructure of the film deposited at $P_T = 50$ Pa, with an anatase content of about 86 vol.%. The film consists of nanoparticles aggregated in clusters with a cauliflower-like morphology. The clusters exhibit a broad size distribution from 40 nm to 337 nm and are composed by numerous particles with a mean size of 40±18 nm. The cauliflower-like morphology is characteristic of the films with prevailing anatase.

At this point we turn our attention to the optical properties of the films, in particular to the optical band gaps. The transmittance spectra in the UV-Vis-NIR region of the as-deposited Co-doped films are shown in Fig. 7. The transmittance increases as $P_T$ increases which can be explained by the phase composition evolution of the samples and the fact that anatase displays a lower absorption coefficient than rutile [24]. Furthermore, interference fringes appear in the optical spectra indicating a low surface roughness of the as-grown films which, however, tends to increase with the total pressure in agreement with the morphological properties.

The occurrence of interference fringes allowed evaluating the thickness of the films, $d$, as follows [25]:

$$d = \frac{M\lambda_1\lambda_2}{2[\lambda_1 n_2(\lambda_2) - \lambda_2 n_1(\lambda_1)]} \tag{3}$$

where $n_1(\lambda_1)$ and $n_2(\lambda_2)$ are the refractive indices at two maxima (or minima) located at $\lambda_1$ and $\lambda_2$, respectively, $\lambda_1 > \lambda_2$ and $M$ is the number of oscillations between the two extremes ($M = 1$ for consecutive maxima or minima). Film thicknesses between 485 nm and 1270 nm were obtained (Table 1). It is worth noting that the thicknesses determined from the RBS measurements agree well with the ones deduced from the optical data. Assuming that intensity losses due to reflection may be neglected, the optical absorption coefficient, $\alpha$, was calculated by the Beer-Lambert law: $\alpha = (1/d)\ln(1/T)$.

The indirect optical band gap energy of the Co:TiO$_2$ films, $E_g$, was estimated by plotting $(\alpha h\upsilon)^{1/2}$ as a function of $h\upsilon$ (Tauc plot), where $h\upsilon$ is the photon energy, and by extrapolating the linear portion of the curve to zero absorption. Fig. 8 shows the Tauc plots obtained for the rutile and anatase films grown at 7 Pa and 60 Pa, respectively. An optical band gap energy of 2.82 eV was determined for the epitaxial rutile Co:TiO$_2$ film while for the pure anatase film the band gap energy was estimated as 3.14 eV. Both values are red shifted compared to the band gaps usually reported for rutile (*ca.* 3.05 eV) and anatase (*ca.* 3.20 eV). One may wonder whether such red shit, $\Delta E_g$, might originate from quantum confinement effects associated with the nanosized crystallites that form the films or might result from the transition metal doping. The structural size effect can be important if the crystallite size is in the range of the Bohr radius of the first excitonic state, which is given by [26]

$$r_B = \frac{m_0 \varepsilon_r}{\mu} a_B \qquad (4)$$

where $m_0$ stands for the electron rest mass, $\varepsilon_r$ for the relative dielectric permittivity, $\mu$ for the reduced electron-hole mass, and $a_B$ for the Bohr radius of the hydrogen atom (5.292×10$^{-11}$ m). If we take the anatase TiO$_2$ electron and hole effective masses to be equal to $m_e = 10m_0$ and $m_h = 0.8m_0$, respectively, and $\varepsilon_r = 12$ [27], we get $r_B = 0.86$ nm. For rutile TiO$_2$, $r_B$ can be estimated to be about 0.26 nm [28]. Since $2r_B$ is much lower than the smallest average crystallite size measured (~8 nm for the film grown at $P_T = 70$ Pa), the quantum confinement effects are certainly very weak and can be ruled out. Moreover, $\Delta E_g$ increases with increasing crystallite size at decreasing $P_T$, as will be seen later, reinforcing the negligible influence of quantum confinement in our samples. Therefore, the red shift of the band gap should result from the Co doping. Indeed, undoped anatase films grown under similar conditions to those used for growing samples G and H show optical band gap energies of 3.19 eV, close to the $E_g$ value of bulk anatase TiO$_2$ referred above. Red shifts are consistent with the introduction of electronic states in the titanium oxide bandgap associated with the 3$d$ electrons of Co$^{2+}$ cations and oxygen defects [29].

Actually, previous band-structure calculations by Park *et al.* [30] and Mo *et al.* [31] have shown that the valence band derives primarily from O 2*p*-levels, the conduction band from the Ti 3*d*-levels, and that the crystal-field split Co 3*d*-levels form localized bands within the original band gap of $TiO_2$ [32]. Consequently, the optical absorption edge transition for the Co-doped $TiO_2$ results from the *n*-type doping of the $TiO_2$ matrix [33]. Similar narrowing of the band gaps were observed in Co-doped anatase $TiO_2$ films [34] and Nd-doped anatase $TiO_2$ nanoparticles [35].

Fig. 9 displays the $E_g$ values as a function of the anatase/rutile content of the films, showing that all Co-doped films exhibit an enhanced absorption in the near-visible region of the electromagnetic spectrum compared to the absorption edge usually reported for undoped $TiO_2$, either for the anatase or rutile phases. Furthermore, the band gap red shift tends to increase with increasing rutile content (decreasing $P_T$) i.e., the rutile-only film presents a higher red shift ($\Delta E_g \approx 0.2$ eV) than the anatase-only film ($\Delta E_g \approx 0.06$ eV). This might be related with the crystal structural features of both polymorphic phases (anatase presents a less dense crystallographic structure than rutile) and thus different crystal field strength splitting of the localized Co 3*d*-states. These localized states form tails of states that extend the bands into the energy gap producing an absorption tail, known as the Urbach tail, which follows the exponential law $\alpha = \alpha_0 \exp(h\nu/E_U)$, where the Urbach energy, $E_U$, is the width of the tail and $\alpha_0$ a constant [36]. The calculation of $E_U$ involves the fitting of ln$\alpha$ in an energy range just below the gap (*cf.* the inset in Fig. 8). The $E_U$ values obtained by this method are given in Table 1 and plotted as a function of the working pressure in the inset of Fig. 9. As the Urbach energy increases, the band gap red shift increases due to possible band-to-tail and tail-to-tail transitions. Similar results were obtained for Ce-doped $TiO_2$ nanoparticles [37] and Mn-doped ZnO films [38].

It is worth noting that the broader range of light absorption of the Co:TiO$_2$ films might favour the photocatalytic activity in the visible region. On the other hand, the variation of $E_g$ with the rutile-to-anatase relative contents suggests the possibility of tailoring the optical band gap by adjusting the phase composition – in our experiments just by changing the working pressure.

## 4. Conclusions

We have shown that pure rutile and anatase Co-doped TiO$_2$ thin films, as well as films with a mixture of both polymorphic phases, can be grown by non-reactive PLD using argon as buffer gas, onto (0001)Al$_2$O$_3$ substrates heated at the low temperature of 310 ºC, by varying only the working pressure in the deposition chamber. Epitaxial rutile films were grown at $P_T = 7$ Pa while pure anatase films were grown at $P_T \geq 60$ Pa. Polycrystalline films with rutile-anatase mixture of phases were obtained for 20 Pa $< P_T <$ 60 Pa, the anatase content increasing with increasing $P_T$. The microstructure of the films strongly depends on their phase composition. Films consisting of pure anatase or for which anatase is the prevalent phase show a porous microstructure with cauliflower-like morphology that could be particularly attractive for applications requiring materials with large specific surface areas. Optical band gaps are red shifted for all samples in comparison with the values usually reported for undoped TiO$_2$, which is related with the introduction of electronic states by the 3d electrons of Co$^{2+}$ cations and oxygen defects, and therefore with *n*-type doping of the TiO$_2$ matrix. The increase of the band gap red shift as the working pressure decreases is consistent with the corresponding increase of the Urbach energy. The narrowed band gaps bring the absorption edge of Co-doped TiO$_2$ films into the near visible region, in relation with the phase composition of the films, which suggests that non-reactive PLD with variation of only one parameter, the working pressure, could provide a method for band gap engineering in the near-visible region.

**Acknowledgements**

This work was supported by the Portuguese Foundation for Science and Technology (FCT), grant no. PTCD/CTM/101033/2008. S.R. acknowledges FCT for a Post-Doctoral grant, reference SFRH/BPD/64390/2009. S.D. acknowledges a Post-Doctoral grant funded by the aforementioned FCT research grant.

**TABLE CAPTION**

**Table 1**. Background gas phase pressure, $P_T$, phase composition, structural and optical parameters of the Co-TiO$_2$ films grown onto (0001)Al$_2$O$_3$ substrates; $<D>_{hkl}$: average crystallite size along the [hkl] direction; $C_A, C_R$: anatase, rutile phase content; $d$: film thickness; $r_{dep}$: deposition rate; $E_g$: optical band gap energy; $E_U$: Urbach energy.

**FIGURE CAPTIONS**

**Figure 1.** RBS spectrum of a Co-doped $TiO_2$ thin film deposited onto $(0001)Al_2O_3$ at $P_T = 10$ Pa. The spectrum was recorded using a 2.0 MeV $He^+$ beam and shows the experimental data and the corresponding simulation from which a Co:Ti ratio of 0.03:0.97 was inferred.

**Figure 2.** a) X-ray diffractogram of a Co-doped $TiO_2$ film deposited onto $(0001)Al_2O_3$ at $P_T = 7$ Pa. Labelled lines were taken from the JCPDS card no. 21-1276 for rutile. The inset shows the rocking curve of the (200) rutile diffraction peak. b) $\varphi$–scan performed around the (301) plane of rutile and around the asymmetrical (108) crystallographic plane of sapphire for the film displayed in Fig. 2a.

**Figure 3.** X-ray diffractograms of Co-doped $TiO_2$ films deposited onto $(0001)Al_2O_3$ at total pressure, $P_T$, ranging from 10 to 70 Pa. Labelled lines were taken from the JCPDS card no. 21-1272 for $TiO_2$ anatase (A) and 21-1276 for rutile (R).

**Figure 4.** Anatase, $C_A$, and rutile, $C_R$, content plotted as a function of total pressure. Data (full squares) were estimated from the XRD patterns in Fig. 2a and Fig. 3, and were fitted by a Gompertz-like sigmoid function (full line). Inset: plots of the derivative of the Gompertz function (full line) and of the deposition rate (open circles) *versus* pressure.

**Figure 5.** Pressure dependence of the average crystallite size, $<D>_{hkl}$, calculated from the *hkl* Bragg peaks displayed in Fig. 2a and Fig. 3: □ - rutile (200) peak; ■ - rutile (101) peak; ● - anatase (101) peak.

**Figure 6.** FEG-SEM micrographs of films prepared at total working pressures of a) $P_T = 7$ Pa (epitaxial rutile film) and b) $P_T = 50$ Pa ($C_A = 0.862$ vol.%). The insets show magnified images. The particle size distribution histograms are displayed under the corresponding micrographs.

**Figure 7.** Optical transmittance spectra of Co-doped $TiO_2$ films grown in argon environment at different total working pressures.

**Figure 8.** $(\alpha h\upsilon)^{1/2}$ *versus* energy curves for the Co-doped $TiO_2$ films grown at $P_T = 7$ Pa

(epitaxial rutile film) and at $P_T = 60$ Pa (anatase-only film). Straight lines represent linear fits to the high-energy portion of the curves. The band gap values of the films are also indicated in the figure. The inset shows the variation of ln$\alpha$ with photon energy allowing the evaluation of the Urbach energy.

**Figure 9.** Optical band gap energy data represented as a function of anatase, $C_A$, and rutile, $C_R$, contents for the Co-doped TiO$_2$ films. The horizontal dash-dot lines indicate the $E_g$ values for the pure anatase and rutile polymorphs. The inset shows the dependence on total pressure for both the band gap, $E_g$, and Urbach, $E_U$, energies.

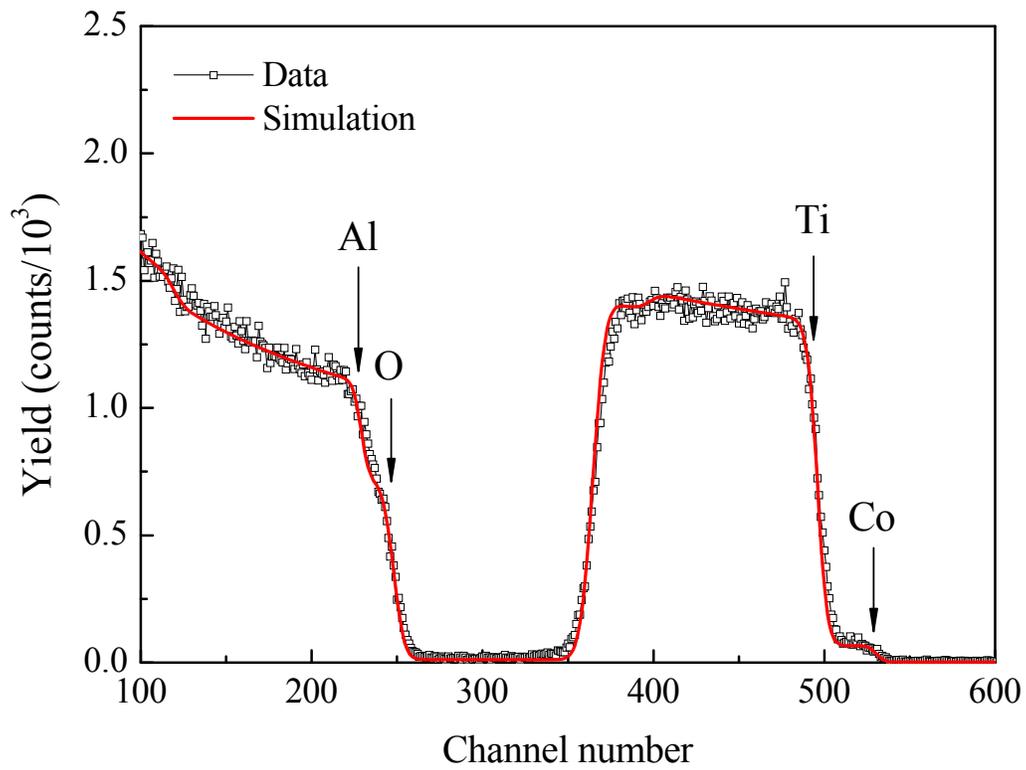

**Figure 1**

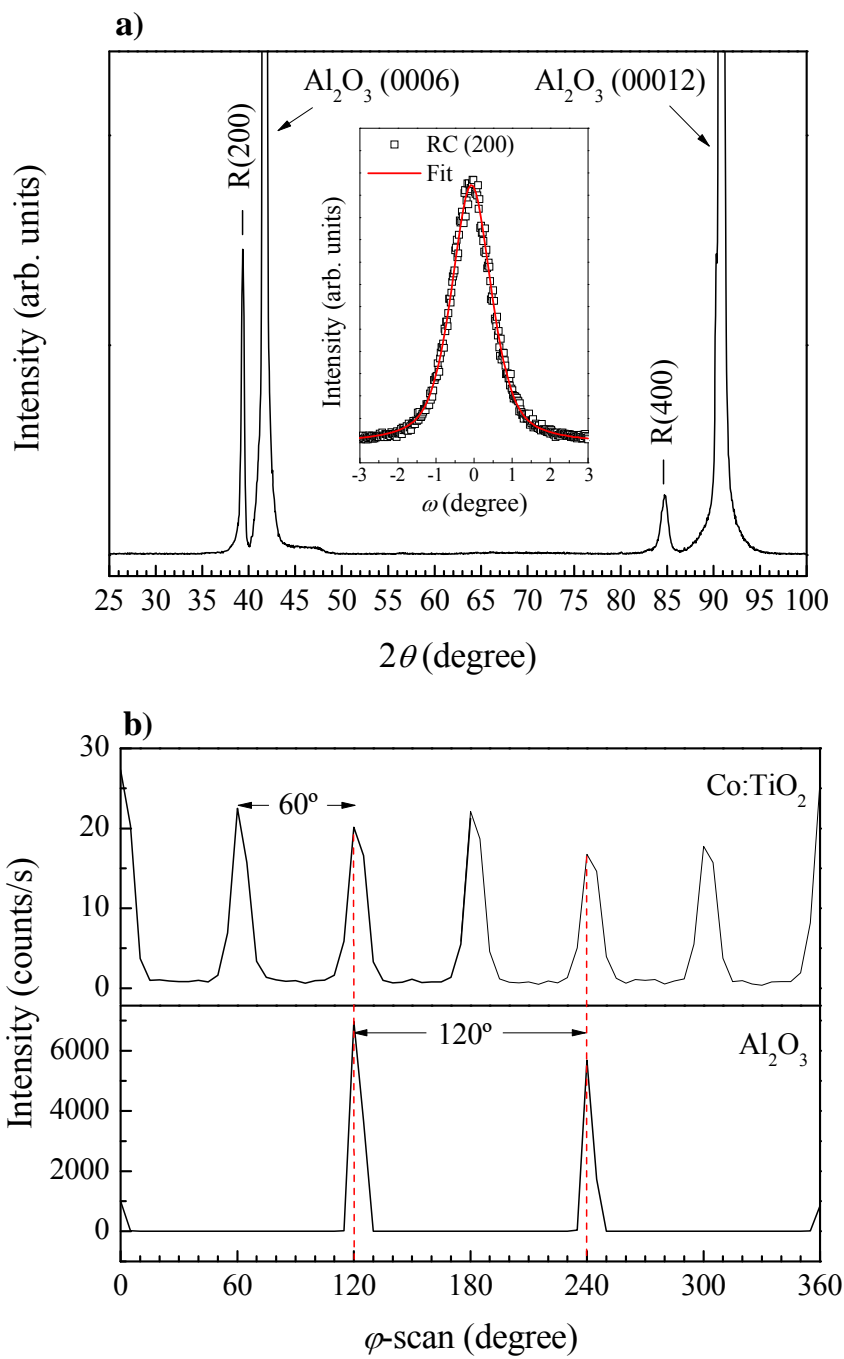

**Figure 2**

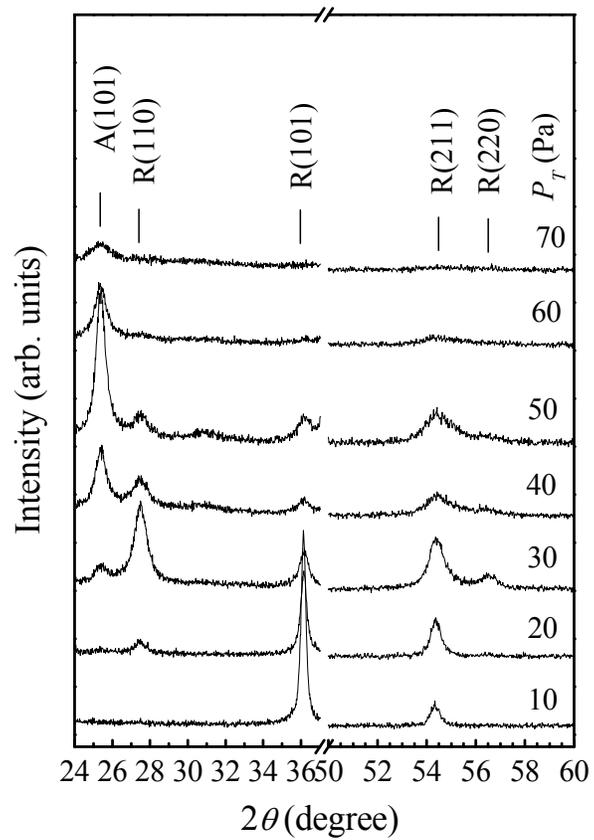

**Figure 3**

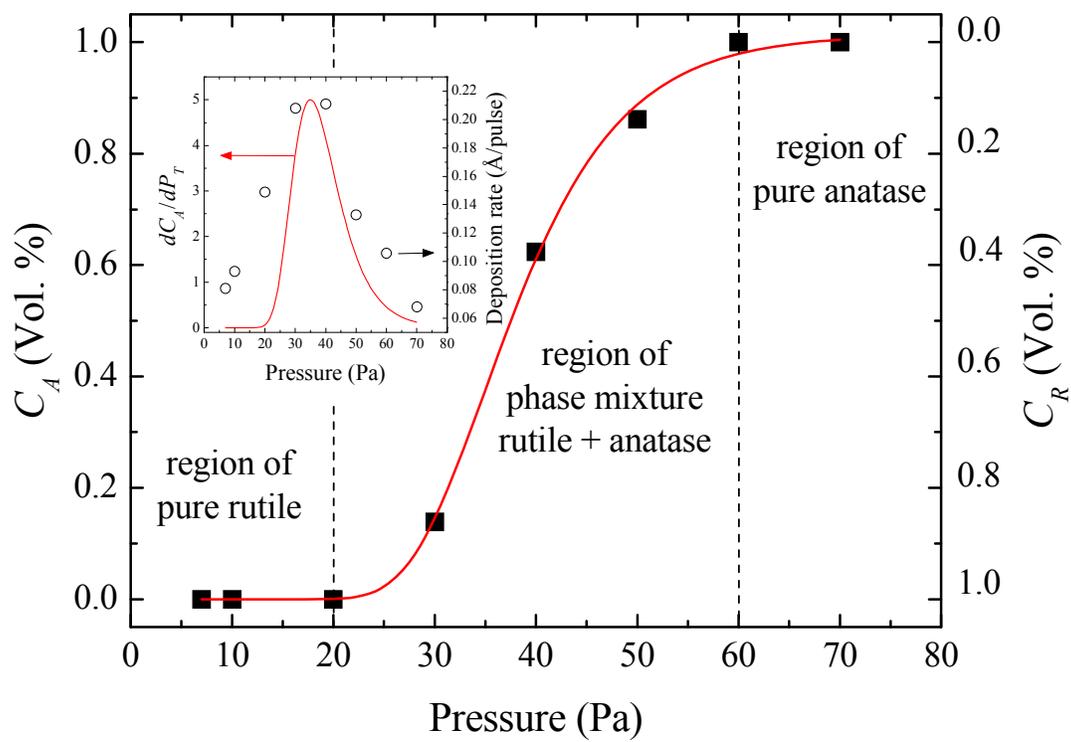

**Figure 4**

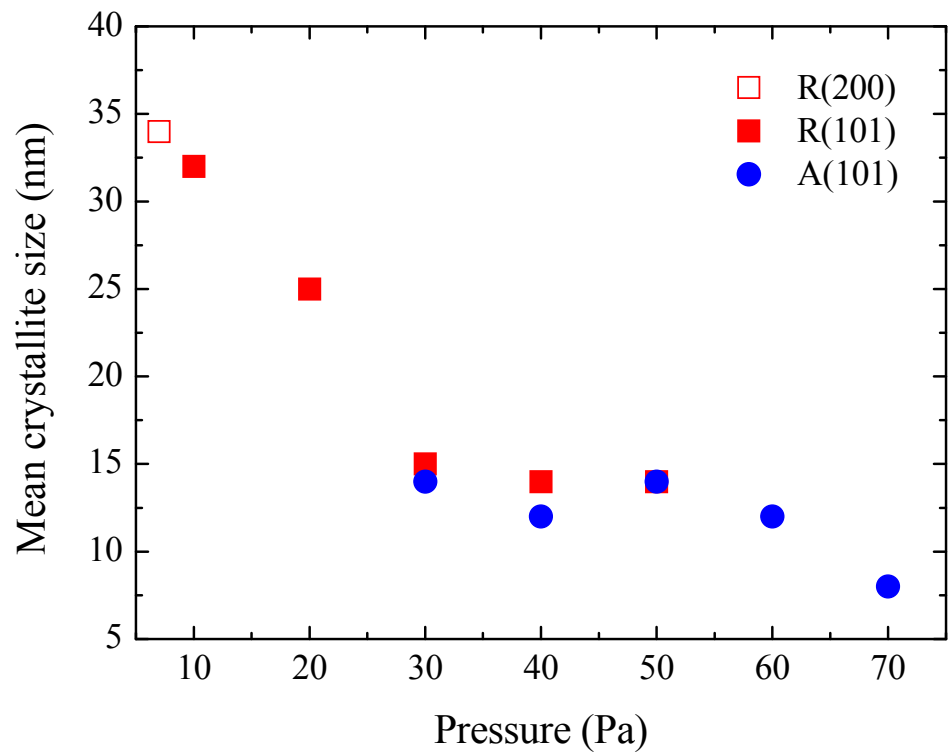

**Figure 5**

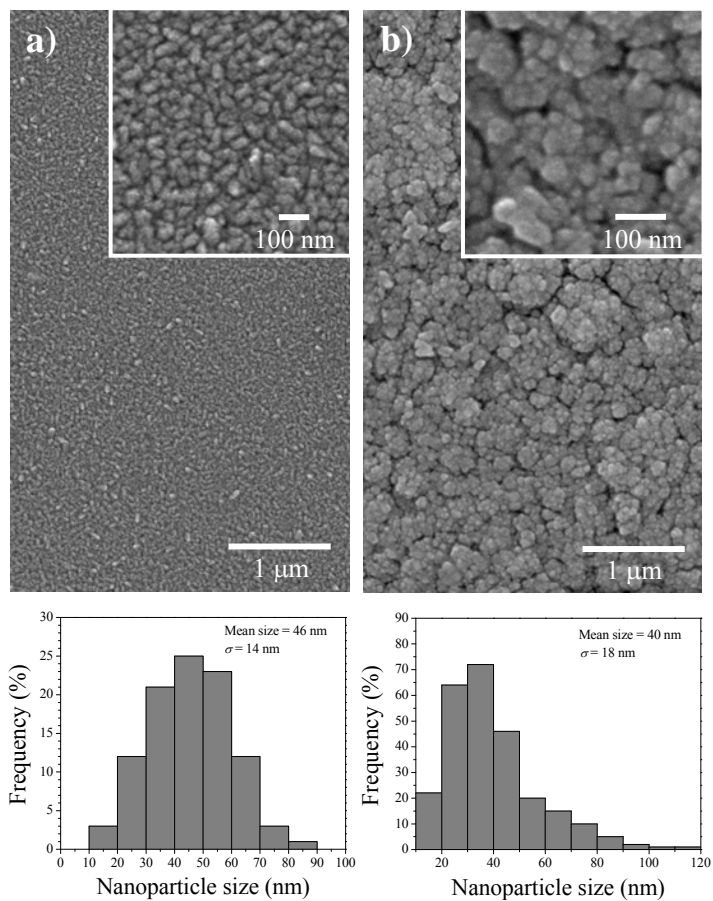

**Figure 6**

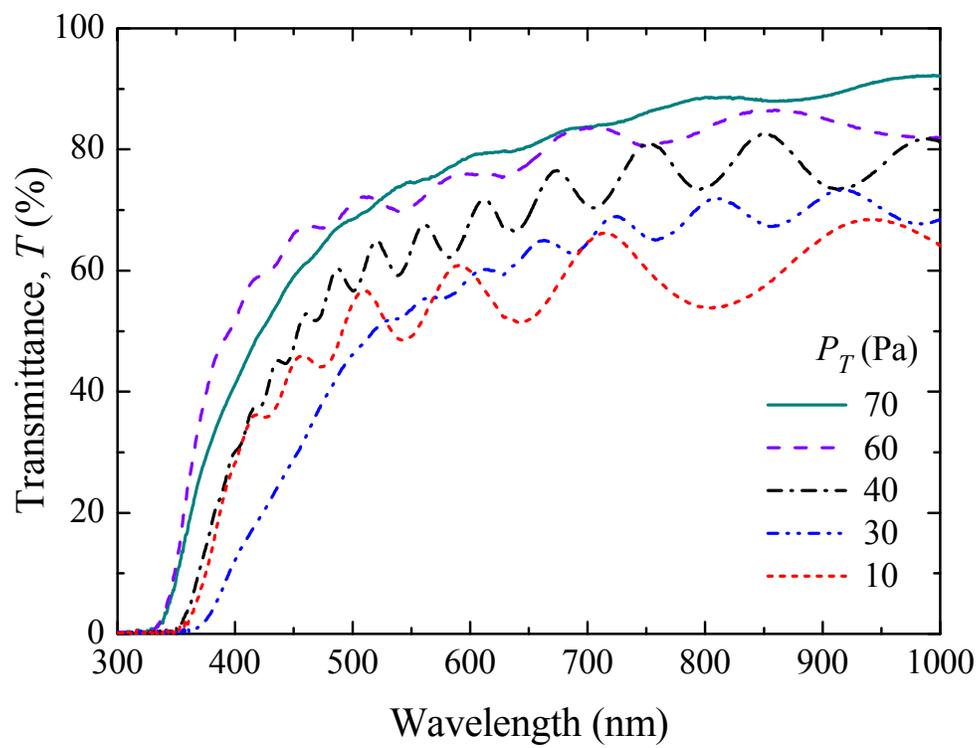

**Figure 7**

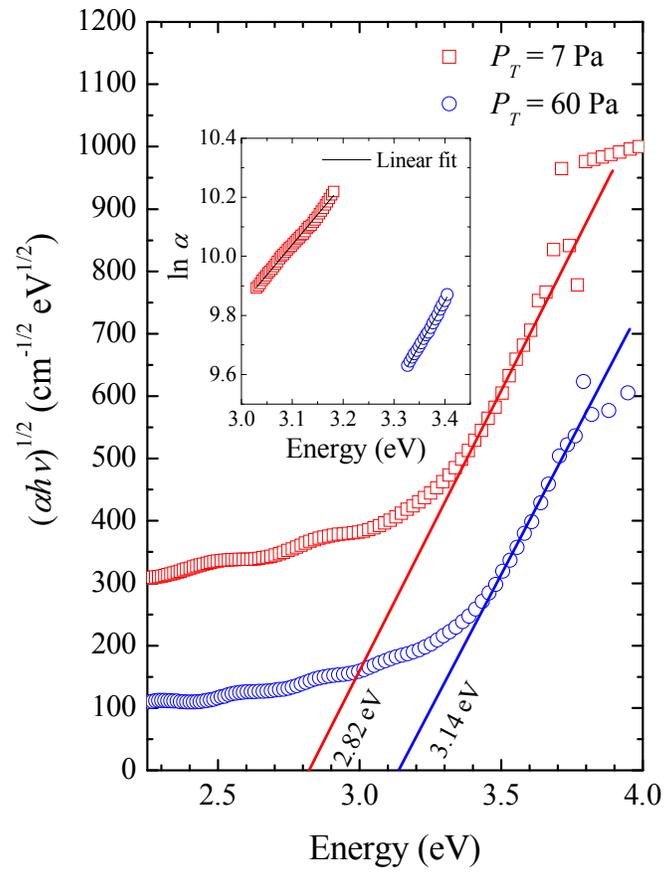

**Figure 8**

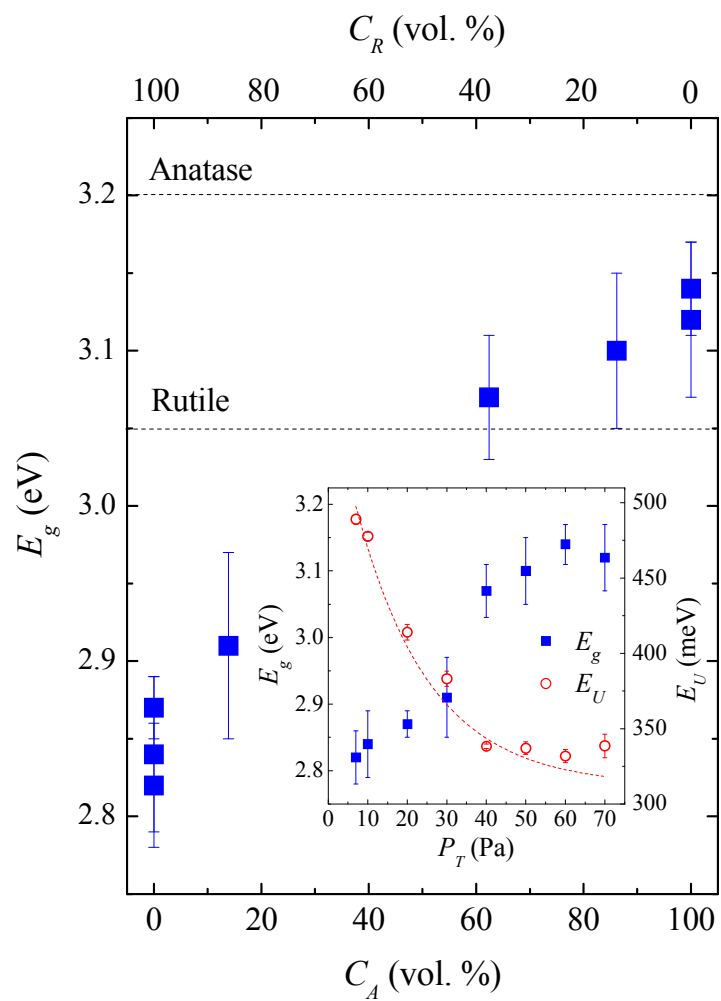

**Figure 9**